\documentclass[conference]{IEEEtran}
\IEEEoverridecommandlockouts

\usepackage[T1]{fontenc}
\usepackage{amsmath,amssymb,amsthm,bm}
\usepackage{graphicx,booktabs,array}
\usepackage{enumitem,float}
\usepackage{cite}
\usepackage[hidelinks]{hyperref}

\floatstyle{ruled}
\newfloat{algorithm}{t}{loa}
\floatname{algorithm}{Algorithm}

\theoremstyle{plain}
\newtheorem{theorem}{Theorem}
\newtheorem{proposition}{Proposition}
\newtheorem{corollary}{Corollary}
\theoremstyle{remark}
\newtheorem{remark}{Remark}
\newenvironment{NewProof}{{\noindent\itshape Proof.}}{\hfill $\blacksquare$\par}

\newcommand{\E}{\mathbb E}
\newcommand{\Var}{\operatorname{Var}}
\newcommand{\Cov}{\operatorname{Cov}}
\newcommand{\argmin}{\operatorname*{arg\,min}}
\newcommand{\one}{\bm 1}
\newcommand{\R}{\mathbb R}

\title{Low-PAPR Rank-2 DFT-s-OFDM \\Waveform for 6G Uplink}
\author{Yewen Cao, Fen Hou, Yulin Shao%
\thanks{Y. Cao and F. Hou are with the State Key Laboratory of Internet of Things for Smart City, University of Macau, Macao, China.}
\thanks{Y. Shao is with the Department of Electrical and Computer Engineering, The University of Hong Kong, Hong Kong, China.}
}

\begin{document}
\maketitle

\begin{abstract}
Rank-2 discrete Fourier transform-spread orthogonal frequency-division multiplexing (DFT-s-OFDM) is attracting renewed
    interest for 6G uplink, offering not only spatial multiplexing gains but also an additional degree of freedom for
    peak-to-average power ratio (PAPR) reduction. We exploit this freedom by selectively exchanging constellation
    symbols between the two layers before DFT spreading, preserving the transmitted information while reshaping the
    resulting waveforms. Building on this idea, we propose Cyclic Layer Assignment with Shared-basis Synthesis (CLASS).
    CLASS employs carefully designed cyclic assignment patterns to generate diverse waveform candidates, while reusing a
    small number of waveform components to avoid synthesizing each candidate separately. We further establish the
    optimality of the cyclic design in maximizing waveform diversity under a minimum-basis constraint when equal group
    sizes are feasible. With only six inverse DFTs, CLASS explores 16 assignments and reduces the 99th-percentile
    worst-layer PAPR by 1.41 dB relative to fixed assignment, outperforming representative selected mapping (SLM) and
    partial transmit sequences (PTS) schemes under the same synthesis budget.
\end{abstract}


\section{Introduction}
The ongoing development of 6G wireless systems has renewed interest in multi-layer discrete Fourier transform-spread
    orthogonal frequency-division multiplexing (DFT-s-OFDM) \cite{MyungLimGoodman2006,ShaoLiew2020} for uplink
    transmission. In particular, the 3GPP has supported uplink DFT-s-OFDM transmission with a maximum rank of two in its
    Release 20 study on 6G radio (TR 38.760-1) \cite{3GPP_TR387601,3GPP_R12605036}.
This development aims to combine the spectral-efficiency benefits of spatial multiplexing with the favorable
    power-efficiency characteristics of DFT-s-OFDM.

These power-efficiency benefits, however, depend on the peak-to-average power ratios (PAPRs) \cite{HanLee2005} of both
    spatial layers, as a high PAPR on either layer may require substantial power-amplifier backoff. Reducing the
    worst-layer PAPR without excessive transmitter complexity is therefore an important design objective for 6G rank-2 DFT-s-OFDM uplink.

A variety of PAPR reduction techniques have been developed for DFT-s-OFDM and its closely related single-carrier
    frequency-division multiple access (SC-FDMA) waveform \cite{MyungLimGoodman2006,cao2025agile,shao2022semantic},
    including signal shaping \cite{JiRenZhang2015,KimYunKimCho2018}, selected mapping (SLM)
    \cite{BaeumlFischerHuber1996,YenMinn2012}, and partial transmit sequences (PTS)
    \cite{MuellerHuber1997,SieglFischer2008}. Extensions to multi-antenna transmission have further explored joint
    waveform selection \cite{FischerSiegl2009,YenMinn2012} and PAPR optimization across transmit branches
    \cite{SieglFischer2008,YenMinn2012}, while cross-antenna signal swapping has been investigated in multiple-input
    multiple-output (MIMO) OFDM systems \cite{GaoZhuNandi2009}. Despite their different approaches, these techniques
    face a fundamental trade-off: achieving greater PAPR reduction generally requires exploring more candidate
    waveforms, which increases the cost of waveform generation and evaluation. For rank-2 DFT-s-OFDM, the challenge is
    to achieve effective joint PAPR reduction without incurring excessive waveform-synthesis complexity.

In this paper, we exploit a distinctive waveform-design freedom inherent in rank-2 DFT-s-OFDM: \emph{inter-layer symbol
    assignment before DFT spreading}. At each DFT input position, a pair of constellation symbols can be assigned to the
    two spatial layers in either order. Such exchanges preserve the transmitted information while producing different
    time-domain waveforms, enabling PAPR reduction through adaptive layer assignment. Building on this insight, we
    propose \emph{Cyclic Layer Assignment with Shared-basis Synthesis} (CLASS), a structured framework that generates
    diverse assignment candidates from a small number of shared waveform bases and selects the one minimizing the
    worst-layer PAPR. CLASS exploits the additional freedom of rank-2 transmission while retaining conventional
    frequency-domain MIMO equalization, with the selected assignment reversed at the receiver.

Turning this idea into an efficient waveform-design method entails three technical challenges. First, the effectiveness
    of assignment selection depends on the diversity among the resulting continuous-time waveforms, which is difficult
    to characterize directly from the assignment patterns. Second, the exponentially large assignment space makes direct
    waveform synthesis and exhaustive search computationally prohibitive. Third, restricting the candidate set to reduce
    synthesis complexity raises the question of how to maximize waveform diversity within a limited synthesis budget.

Our main contributions are summarized as follows:
\begin{itemize}[leftmargin=0.45cm]
    \item We establish the statistical equivalence of fixed layer assignments and derive an exact metric for their
        pairwise waveform diversity. This metric expresses the continuous-time aggregate power differences as a
        kernel-weighted distance between assignment patterns.
    \item We establish the minimum number of differential waveform bases required for a given number of assignment
        candidates and characterize the corresponding assignment structure. This enables exponentially many candidates
        to be synthesized from a logarithmic number of shared bases.
    \item We prove that cyclic grouping maximizes the average pairwise waveform diversity under the minimum-basis
        constraint when equal group sizes are feasible. We further develop an efficient candidate-selection algorithm
        based on incremental waveform updates.
\end{itemize}

Numerical results demonstrate that CLASS achieves a 1.41 dB reduction in the 99th-percentile worst-layer PAPR relative
    to fixed layer assignment and outperforms representative SLM- and PTS-based schemes under the same inverse-DFT synthesis budget.
These results demonstrate that the second spatial layer in rank-2 DFT-s-OFDM offers not only additional transmission
    capacity but also a new degree of freedom that can be systematically exploited for low-PAPR waveform design.

\begin{figure}[!t]
\centering
\includegraphics[width=0.80\columnwidth]{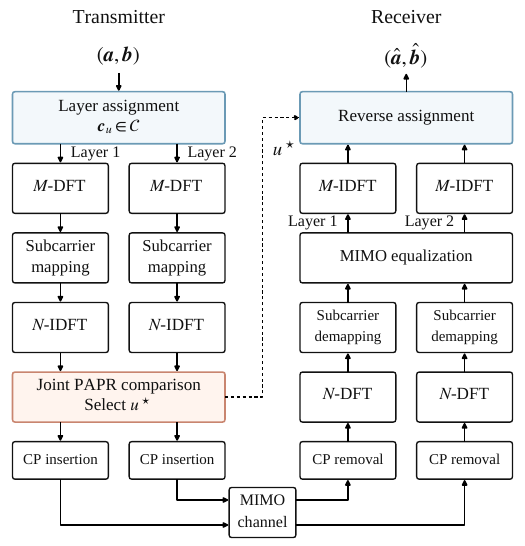}
\caption{Rank-2 DFT-s-OFDM transceiver with CLASS. The same symbol pair is assigned to two layer paths according to
    \(\bm c_u\), and the transmitter selects \(u^\star\) from \(\mathcal C\) according to PAPR. After DFT-s-OFDM
    reception and MIMO equalization, the receiver uses \(u^\star\) to reverse the selected assignment.}
\label{fig:transmitter}
\end{figure}

\section{System Model}
We consider a single-user uplink using rank-2 DFT-s-OFDM with two spatial layers, as illustrated in
    Fig.~\ref{fig:transmitter}. For each layer \(\ell\in\{0,1\}\), an \(M\)-point DFT is followed by mapping to \(M\)
    contiguous subcarriers and an \(N\)-point inverse DFT (IDFT), where \(M\le N\) \cite{MyungLimGoodman2006,park2011}.
    We consider non-coherent rank-2 transmission, for which the two layer waveforms are not combined by spatial
    precoding \cite{3GPP_TR387601}.

\subsection{Layer Assignment}
For one DFT-s-OFDM block, we arrange the \(2M\) constellation symbols into two length-\(M\) sequences \(\bm
    a=[a[0],\ldots,a[M-1]]^{\mathsf T}\) and \(\bm b=[b[0],\ldots,b[M-1]]^{\mathsf T}\). Their entries are independent
    and identically distributed with unit average energy. At DFT input index \(m=0,\ldots,M-1\), the pair
    \((a[m],b[m])\) can be assigned to the two layers in either order.

Let \(\bm c=[c[0],\ldots,c[M-1]]^{\mathsf T}\in\{-1,+1\}^M\) specify these choices. The two layer inputs are
\begin{equation}
\bigl(x_{0,\bm c}[m],x_{1,\bm c}[m]\bigr)=
\begin{cases}
(a[m],b[m]), & c[m]=+1,\\
(b[m],a[m]), & c[m]=-1.
\end{cases}
\label{eq:mapper}
\end{equation}
Thus each entry of \(\bm c\) either keeps or exchanges one symbol pair. CLASS considers \(U\) assignment vectors
    collected in \(\mathcal C=\{\bm c_0,\ldots,\bm c_{U-1}\}\).

\subsection{Continuous-Time Waveform}
For an assignment \(\bm c\), the continuous-time baseband signal of layer \(\ell\) is
\begin{equation}
s_{\ell,\bm c}(t)=\frac1{\sqrt M}\sum_{m=0}^{M-1}
 x_{\ell,\bm c}[m]\phi_m(t),\qquad 0\le t<T,
\label{eq:waveform}
\end{equation}
where \(\phi_m(t)=M^{-1/2}\sum_{k=0}^{M-1}e^{j2\pi k(t/T-m/M)}\) and \(T\) is the useful DFT-s-OFDM symbol duration. Here the occupied subcarriers are indexed by \(k=0,\ldots,M-1\). The pulse \(\phi_m(t)\) describes the continuous waveform contribution of the symbol at DFT input index \(m\). Its cardinal values satisfy \(\phi_m(m'T/M)=\sqrt M\,\mathbf1_{\{m=m'\}}\), so \eqref{eq:waveform} reconstructs the assigned constellation symbols at those instants.

We evaluate an assignment by the larger PAPR of the two layer waveforms \cite{venkataraman2006}:
\begin{equation}
J(\bm c)=\max_{\ell\in\{0,1\}}
\frac{\max_{0\le t<T}|s_{\ell,\bm c}(t)|^2}
{P_{\ell,\bm c}},
\label{eq:objective}
\end{equation}
where \(P_{\ell,\bm c}=T^{-1}\int_0^T |s_{\ell,\bm c}(t)|^2\,dt=\|\bm x_{\ell,\bm c}\|^2/M\) is the average power of layer \(\ell\) over one useful symbol. The transmitter selects
\[
u^\star\in\argmin_{0\le u<U}J(\bm c_u).
\]
A cyclic prefix (CP) is then appended before transmission. The selected index \(u^\star\), represented by
    \(\lceil\log_2U\rceil\) bits, is conveyed to the receiver. After CP removal, the receiver applies an \(N\)-point
    DFT, selects the \(M\) occupied subcarriers, and performs frequency-domain MIMO equalization followed by an
    \(M\)-point IDFT on each layer \cite{prasad2009}. This yields estimates of \(\bm x_{0,\bm c_{u^\star}}\) and \(\bm
    x_{1,\bm c_{u^\star}}\). Using the conveyed index \(u^\star\), the receiver reverses \eqref{eq:mapper} to recover
    \(\bm a\) and \(\bm b\).

\begin{remark}[Unchanged MIMO Equalization]
The proposed assignment only exchanges the DFT input symbols between the two spatial layers and does not alter the
    occupied subcarriers. Therefore, under a CP-based linear time-invariant (LTI) channel model, the conventional
    frequency-domain MIMO equalization remains unchanged. Each occupied subcarrier is still equalized independently
    using its corresponding MIMO channel matrix. The selected index \(u^\star\) is used only after the two \(M\)-point
    IDFTs to reverse the exchanges in \eqref{eq:mapper}.
\end{remark}

\section{Candidate Diversity}
The additional layer assignment freedom in rank-2 DFT-s-OFDM creates a choice among waveform realizations for each data
    block. The PAPR statistics of a fixed assignment characterize one alternative in isolation, while the benefit of
    selection depends on how the alternatives behave together. We examine these two aspects through fixed-assignment
    statistics and pairwise waveform diversity.

\subsection{Origin of Waveform Diversity}
For a fixed \(\bm c\), \eqref{eq:mapper} only determines which member of each independent symbol pair enters each layer.
    We therefore begin by establishing whether the choice of a fixed assignment changes the PAPR statistics.

\begin{proposition}[Statistical Equivalence of Fixed Assignments]
\label{prop:statistical-equivalence}
For any deterministic vector \(\bm c\), the pair \((\bm x_{0,\bm c},\bm x_{1,\bm c})\) has the same joint distribution
    as \((\bm a,\bm b)\). Consequently, \(J(\bm c)\overset{d}=J(\one)\).
\end{proposition}
\begin{NewProof}
For each \(m\), \eqref{eq:mapper} either keeps or exchanges the independent and identically distributed pair
    \((a[m],b[m])\), so the pair distribution is unchanged. Independence across the input indices then preserves the
    joint distribution of the two sequences. The waveform mapping in \eqref{eq:waveform} and the PAPR metric in
    \eqref{eq:objective} are deterministic, which proves the result.
\end{NewProof}

Proposition~\ref{prop:statistical-equivalence} establishes that no fixed assignment is statistically preferred. Hence
    any PAPR reduction from \eqref{eq:objective} comes from choosing among the different waveforms produced by the same
    data block. Let \(F_1(\gamma)\) denote the cumulative distribution function (CDF) of the PAPR of either layer under
    a fixed assignment. Since the two layer PAPRs are independent and identically distributed, the fixed rank-2 PAPR CDF
    is \(F_{\rm fixed}(\gamma)=F_1^2(\gamma)\).

To identify what changes across these statistically equivalent assignments, define the common and differential symbol
    components \(\bm p=(\bm a+\bm b)/2\) and \(\bm q=(\bm a-\bm b)/2\). Their corresponding waveforms are
\[
\begin{aligned}
y(t)&=\frac1{\sqrt M}\sum_{m=0}^{M-1}p[m]\phi_m(t),\\
z_{\bm c}(t)&=\frac1{\sqrt M}\sum_{m=0}^{M-1}c[m]q[m]\phi_m(t).
\end{aligned}
\]
Since \(\bm x_{0,\bm c}=\bm p+\bm c\odot\bm q\) and \(\bm x_{1,\bm c}=\bm p-\bm c\odot\bm q\), where \(\odot\) denotes
    entrywise multiplication, linearity gives
\begin{equation}
s_{0,\bm c}=y+z_{\bm c},\qquad
s_{1,\bm c}=y-z_{\bm c}.
\label{eq:decomposition}
\end{equation}
The common waveform \(y\) is shared by all candidates, while the assignment acts only through \(z_{\bm c}\). Changing
    \(\bm c\) therefore changes how the differential waveform combines with \(y\) on the two layers. A global sign
    reversal gives \(z_{-\bm c}=-z_{\bm c}\) and only exchanges the two layer waveforms, so \(J(-\bm c)=J(\bm c)\). Thus
    \(\bm c\) and \(-\bm c\) represent the same PAPR candidate.

\subsection{Pairwise Waveform Diversity}
The decomposition above identifies the source of waveform variation, but designing \(\mathcal C\) requires a measure of
    how differently two candidates respond to the same data block. If two alternatives produce nearly the same peak
    behavior, selecting between them provides little additional opportunity for PAPR reduction. SLM studies have related
    candidate correlation to PAPR performance \cite{woo2015}. We therefore quantify the pairwise waveform separation
    among the statistically equivalent assignments considered here. For candidates \(u\) and \(v\), define the aggregate
    instantaneous power \(T_u(t)=|s_{0,\bm c_u}(t)|^2+|s_{1,\bm c_u}(t)|^2\). From \eqref{eq:decomposition},
    \(T_u(t)=2|y(t)|^2+2|z_{\bm c_u}(t)|^2\), so the common term cancels in \(T_u(t)-T_v(t)\). We measure their pairwise
    waveform diversity by
\begin{equation}
D_{uv}\triangleq\frac1{2T}\int_0^T
\E\!\left[(T_u(t)-T_v(t))^2\right]dt.
\label{eq:pairwise-diversity}
\end{equation}

To express \(D_{uv}\) in terms of the assignments, let \(h_{uv}[m]\triangleq c_u[m]c_v[m]\) and \(\bm
    h_{uv}=[h_{uv}[0],\ldots,h_{uv}[M-1]]^{\mathsf T}\). Thus \(h_{uv}[m]=+1\) when the two candidates make the same
    assignment choice at index \(m\), and \(-1\) otherwise. The overlap between the interpolation powers associated with
    indices \(m\) and \(m'\), averaged over one useful symbol, is
\[
K_{m,m'}\triangleq
\frac1T\int_0^T|\phi_m(t)|^2|\phi_{m'}(t)|^2\,dt
=K_{(m-m')\bmod M},
\]
where
\begin{equation}
K_\Delta=
\begin{cases}
(2M^2+1)/(3M), & \Delta=0,\\
1/[M\sin^2(\pi\Delta/M)], & 1\le\Delta<M.
\end{cases}
\label{eq:kernel}
\end{equation}
Here \(\Delta\) is the circular separation between the two DFT input indices. Hence \(K_{m,m'}\) weights the influence
    of their relative assignment choices on the continuous-time power. Let \(\bm K=[K_{m,m'}]_{m,m'=0}^{M-1}\).

\begin{theorem}[Pairwise Diversity Characterization]
\label{thm:candidate-diversity}
Let \(A\) denote a generic constellation symbol satisfying \(\E[A]=0\), \(\E[A^2]=0\), and \(\E[|A|^4]<\infty\). Then
\begin{equation}
\begin{aligned}
D_{uv}
&=1-\frac{\bm h_{uv}^{\mathsf T}\bm K\bm h_{uv}}{M^2}\\
&=\frac{4}{M^2}\sum_{0\le m<m'<M}
K_{m,m'}\mathbf1_{\{h_{uv}[m]\ne h_{uv}[m']\}}.
\end{aligned}
\label{eq:diversity}
\end{equation}
Moreover, \(D_{uv}=0\) if and only if \(\bm c_v=\pm\bm c_u\).
\end{theorem}
\begin{NewProof}
Since \(\E|q[m]|^2=1/2\) and \(\sum_m|\phi_m(t)|^2=M\), the normalized correlation of \(z_{\bm c_u}(t)\) and \(z_{\bm c_v}(t)\) is
\[
\rho_{uv}(t)=\frac1M\sum_{m=0}^{M-1}
 h_{uv}[m]|\phi_m(t)|^2.
\]
Expanding the pulse powers and integrating gives
\[
\frac1T\int_0^T|\rho_{uv}(t)|^2dt
=\frac{\bm h_{uv}^{\mathsf T}\bm K\bm h_{uv}}{M^2}.
\]

From \eqref{eq:decomposition}, \(T_u(t)=2|y(t)|^2+2|z_{\bm c_u}(t)|^2\). Expanding \(\Var(T_u(t))\) and
    \(\Cov(T_u(t),T_v(t))\) over the independent input indices shows that the fourth moment terms common to the two
    expressions cancel. The terms involving \(\E[A^2]\) vanish under the stated conditions, leaving
\[
\Var(T_u(t))-\Cov(T_u(t),T_v(t))=1-|\rho_{uv}(t)|^2.
\]
By Proposition~\ref{prop:statistical-equivalence}, \(T_u(t)\) and \(T_v(t)\) have the same mean and variance. Therefore
\[
\frac12\E[(T_u(t)-T_v(t))^2]
=\Var(T_u(t))-\Cov(T_u(t),T_v(t)).
\]
Combining these relations with \eqref{eq:pairwise-diversity} proves the first equality in \eqref{eq:diversity}.

The entries of \(\bm K\) sum to \(M^2\). Also, \(h_{uv}[m]h_{uv}[m']\) equals \(+1\) when the two relative signs agree
    and \(-1\) when they differ. Substituting these signs into the first expression in \eqref{eq:diversity} gives the
    second equality. All off-diagonal weights are positive, so \(D_{uv}=0\) exactly when all entries of \(\bm h_{uv}\)
    are equal, which is equivalent to \(\bm c_v=\pm\bm c_u\).
\end{NewProof}

The moment conditions in Theorem~\ref{thm:candidate-diversity} hold for square quadrature amplitude modulation (QAM)
    constellations. The theorem converts the continuous-time, data-averaged definition in \eqref{eq:pairwise-diversity}
    into a deterministic quantity that depends only on the two assignments and the fixed kernel \(\bm K\). Thus
    candidate diversity can be evaluated before transmission and used to organize \(\mathcal C\). The weighted form in
    \eqref{eq:diversity} further shows that Hamming distance alone does not capture waveform separation, because
    disagreements are weighted by the circular separation of the corresponding DFT input indices. The condition
    \(D_{uv}=0\) if and only if \(\bm c_v=\pm\bm c_u\) also shows that zero diversity occurs exactly between equivalent PAPR candidates.

To design the full candidate set, define the average pairwise diversity as \(\bar D\triangleq[U(U-1)]^{-1}\sum_{u\ne
    v}D_{uv}\) for \(U\ge2\). For each input index, collect its signs across the candidates as \(\bm
    v_m=[c_0[m],\ldots,c_{U-1}[m]]^{\mathsf T}\). Summing the first expression in \eqref{eq:diversity} over all candidate pairs gives
\begin{equation}
\bar D=\frac{U}{U-1}
\left[
1-\frac{1}{M^2U^2}\sum_{m,m'=0}^{M-1}
K_{m,m'}(\bm v_m^{\mathsf T}\bm v_{m'})^2
\right].
\label{eq:general-set-diversity}
\end{equation}
Equation~\eqref{eq:general-set-diversity} provides the criterion for arranging \(U\) assignments in Sec.~IV.

\section{CLASS Design}
A large set of candidate waveforms does not necessarily require synthesizing each candidate independently. The
    decomposition in \eqref{eq:decomposition} reveals how different assignments can share waveform components, while
    their diversity depends on how the assignments are organized. CLASS exploits this structure by seeking the minimum
    number of differential bases for a prescribed candidate count \(U\) and maximizing average diversity within that representation.

\subsection{Minimum Differential Basis Count}
The candidate vectors may lie in a low-dimensional subspace, allowing their differential waveforms to share a smaller
    set of basis waveforms. We therefore characterize a candidate set \(\mathcal C\) by \(r(\mathcal
    C)\triangleq\dim\operatorname{span}_{\R}\mathcal C\). Every candidate \(\bm c_u\) can be expressed using a basis of
    this \(r\)-dimensional span. Hence \(\bm c_u\odot\bm q\), and therefore its differential waveform, can be
    synthesized from the corresponding \(r\) basis waveforms using the same coefficients. Since \(\bm c\) and \(-\bm c\)
    represent the same PAPR candidate, we retain only one vector from each pair \(\{\bm c,-\bm c\}\) and use \(\one\) as
    the reference candidate.

This representation uses at most \(r(\mathcal C)\) differential basis waveforms. To support arbitrary data blocks, this
    count cannot in general be reduced. In particular, when all entries of \(\bm q\) are nonzero, elementwise
    multiplication by \(\bm q\) preserves the span dimension, while the interpolation property in \eqref{eq:waveform}
    makes the mapping from the DFT input to the continuous-time waveform injective. Therefore, \(r(\mathcal C)\) is the
    minimum number of differential basis waveforms required to synthesize every candidate exactly. The following theorem
    determines this minimum and characterizes the candidate structure when the bound is met with equality.

\begin{theorem}[Minimum Differential Basis Count]
\label{thm:min-dimension}
Let \(\mathcal C\subset\{-1,+1\}^{M}\) contain \(U\) vectors, no two of which are equal up to a global sign, and let
    \(r=r(\mathcal C)\). Then
\begin{equation}
U\le 2^{r-1},\qquad r\ge 1+\lceil\log_2 U\rceil.
\label{eq:dimension}
\end{equation}
The lower bound is achievable whenever \(1+\lceil\log_2 U\rceil\le M\). When \(U=2^{r-1}\) and \(\one\in\mathcal C\),
    every such candidate set has the following structure: the \(M\) coordinates can be partitioned into \(r\) nonempty
    groups \(\mathcal I_0,\ldots,\mathcal I_{r-1}\) such that every candidate is constant within each group. Across the
    retained candidates and their complements, all \(2^r\) combinations of the group signs occur. Conversely, a set
    constructed in this way has \(r(\mathcal C)=r\).
\end{theorem}
\begin{NewProof}
Let \(\mathcal S=\operatorname{span}_{\R}\mathcal C\), with dimension \(r\). Form an \(M\times r\) matrix whose columns
    are a basis of \(\mathcal S\). Its row rank is \(r\), so some \(r\) rows are linearly independent. Selecting the
    corresponding coordinate indices makes the coordinate projection from \(\mathcal S\) into \(\R^r\) injective. Every
    hypercube vertex in \(\mathcal S\) therefore maps to a distinct sign vector in \(\{-1,+1\}^r\), which gives
    \(|\mathcal S\cap\{-1,+1\}^M|\le2^r\). The \(U\) retained vectors and their complements are \(2U\) distinct vertices
    in \(\mathcal S\). Hence \(2U\le2^r\), proving \eqref{eq:dimension}.

For achievability, let \(r=1+\lceil\log_2U\rceil\) and partition the \(M\) coordinates into \(r\) nonempty groups. The
    grouped family contains \(2^{r-1}\ge U\) complement pairs. Select the all-positive group pattern together with
    \(r-1\) patterns that reverse one of groups \(1,\ldots,r-1\); these \(r\) representatives are linearly independent.
    Additional representatives can be added arbitrarily until \(U\) are selected. The resulting candidate set has span
    dimension \(r\), so the lower bound is attained.

Now suppose \(2U=2^r\). All \(2^r\) projected sign vectors occur. Any remaining coordinate is a linear form \(\bm
    w^{\mathsf T}\bm e\) of the projected signs \(\bm e\), and it must lie in \(\{-1,+1\}\) for every \(\bm
    e\in\{-1,+1\}^r\). Squaring gives \(1=\sum_iw_i^2+2\sum_{i<k}w_iw_ke_ie_k\). The constant function and the products
    \(e_i e_k\) are linearly independent on the sign cube. Hence \(\sum_iw_i^2=1\) and \(w_iw_k=0\) for \(i\ne k\).
    Exactly one coefficient is nonzero, with value \(+1\) or \(-1\). Every coordinate therefore copies one projected
    sign, possibly with a fixed sign reversal. Because \(\one\in\mathcal C\), orient the projected signs so that this
    vertex has all positive entries. The fixed reversals then disappear. Coordinates copying the same projected sign
    form one nonempty group. Conversely, the indicators of \(r\) nonempty disjoint groups are linearly independent, and
    the complete grouped family spans them.
\end{NewProof}

For the power-of-two candidate counts used below, the minimum is \(r=1+\log_2U\). Thus the number of available
    candidates grows exponentially with the number of differential bases. Let \(\beta_{g,u}\in\{-1,+1\}\) be the sign of
    candidate \(u\) on group \(\mathcal I_g\), and define its differential basis waveform as
    \(z_g(t)=M^{-1/2}\sum_{m\in\mathcal I_g}q[m]\phi_m(t)\). Then \(z_{\bm c_u}(t)=\sum_{g=0}^{r-1}\beta_{g,u}z_g(t)\),
    so all differential waveforms are synthesized from the same \(r\) bases. Together with the common waveform, these
    bases generate all layer waveform candidates. Choose the representative in each pair such that \(\beta_{0,u}=1\).
    For \(U=2^{r-1}\), the remaining \(r-1\) group signs span all possible combinations. Hence the vectors \(\bm v_m\)
    in \eqref{eq:general-set-diversity} have inner product \(U\) when \(m\) and \(m'\) belong to the same group and zero
    otherwise. Candidate design therefore reduces to partitioning the \(M\) DFT input indices into \(r\) groups, which
    we optimize next using \(\bar D\).

\subsection{Optimal Cyclic Grouping}
With the minimum basis count established, the remaining design variable is the partition of the \(M\) DFT input indices
    into \(r\) groups. For \(U=2^{r-1}\ge2\), substituting the group sign structure into \eqref{eq:general-set-diversity} gives
\begin{equation}
\bar D=\frac{U}{U-1}\left[
1-\frac1{M^2}\sum_{g=0}^{r-1}
\sum_{m,m'\in\mathcal I_g}K_{m,m'}
\right].
\label{eq:partitionobjective}
\end{equation}
The objective therefore favors partitions in which indices sharing the same group sign have little interpolation
    overlap. Since nearby indices on the circular DFT input grid have larger overlap, this suggests spreading the
    indices of each group as evenly as possible. Define the cyclic groups by
\begin{equation}
\mathcal I_g=\{m:0\le m<M,\ m\bmod r=g\},
\label{eq:cyclic}
\end{equation}
for \(g=0,\ldots,r-1\). Their sizes are \(\lfloor M/r\rfloor\) or \(\lceil M/r\rceil\). When \(M=rG\), each group contains \(G\) equally spaced indices. The following theorem shows that this structure is optimal for \(\bar D\).

\begin{theorem}[Optimal Cyclic Grouping]
\label{thm:cyclic-grouping}
Suppose \(U=2^{r-1}\ge2\) and \(M=rG\). Among all candidate sets \(\mathcal C\) that contain \(\one\) and satisfy
    \(r(\mathcal C)=r\), the cyclic partition in \eqref{eq:cyclic} maximizes \(\bar D\), up to relabeling of the groups. The maximum is
\begin{equation}
\bar D^\star(U,r)=\frac{U(r^2-1)}{3r^2(U-1)}.
\label{eq:cyclicdiversity}
\end{equation}
\end{theorem}
\begin{NewProof}
Let \(S=\sum_g\sum_{m,m'\in\mathcal I_g}K_{m,m'}\). By \eqref{eq:partitionobjective}, maximizing \(\bar D\) is
    equivalent to minimizing \(S\). Let the group sizes be \(M_0,\ldots,M_{r-1}\), with \(\sum_gM_g=M\).

Consider one group with \(M_g\) circularly ordered input indices at angles \(\theta_0,\ldots,\theta_{M_g-1}\), with
    \(\theta_{i+M_g}=\theta_i+2\pi\). For each \(d=1,\ldots,M_g-1\), the \(M_g\) forward arc lengths
    \(\theta_{i+d}-\theta_i\) have mean \(2\pi d/M_g\). Since \(\csc^2(\theta/2)\) is strictly convex on \((0,2\pi)\),
    Jensen's inequality gives
\[
\sum_{i=0}^{M_g-1}\csc^2\!\left(\frac{\theta_{i+d}-\theta_i}{2}\right)
\ge M_g\csc^2\!\left(\frac{\pi d}{M_g}\right).
\]
Summing over \(d\) counts every ordered distinct pair in the group \cite{shao2024theory}. Using
    \(\sum_{d=1}^{M_g-1}\csc^2(\pi d/M_g)=(M_g^2-1)/3\), the off-diagonal overlap of this group is at least
    \((M_g^3-M_g)/(3M)\), with equality when its indices are equally spaced.

The diagonal terms contribute \(MK_0\), so
\[
S\ge MK_0+\frac1{3M}\sum_{g=0}^{r-1}(M_g^3-M_g).
\]
Since \(\sum_gM_g=M\), convexity of \(x^3\) gives \(\sum_gM_g^3\ge M^3/r^2\), with equality when every group has size \(M/r=G\). Therefore
\[
S\ge M^2\left(\frac23+\frac1{3r^2}\right).
\]
The cyclic partition in \eqref{eq:cyclic} has equal group sizes and equally spaced indices within every group, so it
    attains equality in both steps. The equality conditions also show that any optimizer on the \(M\)-point grid is a
    relabeling of these residue classes. Substitution into \eqref{eq:partitionobjective} proves \eqref{eq:cyclicdiversity}.
\end{NewProof}

Theorem~\ref{thm:cyclic-grouping} establishes the largest diversity under the minimum basis structure. The following
    comparison quantifies how much diversity the cyclic design retains relative to an unrestricted set of \(U\) candidates.

\begin{corollary}[Diversity Guarantee]
\label{cor:unrestricted-guarantee}
Under the assumptions of Theorem~\ref{thm:cyclic-grouping}, let \(\bar D_{\rm opt}(U,M)\) denote the largest \(\bar D\)
    among all \(U\) distinct PAPR candidates with no restriction on \(r\). Then \(\bar D_{\rm
    opt}(U,M)\le[U/(U-1)](M^2-1)/(3M^2)\), and the cyclic design satisfies
\begin{equation}
\frac{\bar D^\star(U,r)}{\bar D_{\rm opt}(U,M)}
\ge\frac{1-r^{-2}}{1-M^{-2}}.
\label{eq:diversity-guarantee}
\end{equation}
\end{corollary}
\begin{NewProof}
Every term in the sum of \eqref{eq:general-set-diversity} is nonnegative and \(\bm v_m^{\mathsf T}\bm v_m=U\). Retaining
    only the diagonal terms gives \(\sum_{m,m'}K_{m,m'}(\bm v_m^{\mathsf T}\bm v_{m'})^2\ge MK_0U^2\). Using
    \(K_0=(2M^2+1)/(3M)\) in \eqref{eq:general-set-diversity} gives the stated upper bound. Dividing
    \eqref{eq:cyclicdiversity} by this upper bound gives \eqref{eq:diversity-guarantee}.
\end{NewProof}

For \(U=8\), the minimum basis count is \(r=4\), and the guarantee in \eqref{eq:diversity-guarantee} approaches
    \(93.75\%\) as \(M\) grows. Thus four differential bases retain at least \(93.75\%\) of the maximum average
    diversity permitted by an unrestricted design with eight candidates for large \(M\).

\subsection{Efficient Synthesis}
\label{sec:implementation}

\begin{algorithm}[!t]
\caption{CLASS candidate selection}
\label{alg:assignment-selection}
\footnotesize \textbf{Input:} constellation symbols \(\bm a,\bm b\), group count \(r\)\\ \textbf{Output:} selected index \(u^\star\) and layer waveforms
\begin{enumerate}[label=\arabic*:,leftmargin=0.36cm,itemsep=0pt,topsep=1pt,parsep=0pt,partopsep=0pt]
\item Form \(\bm p=(\bm a+\bm b)/2\), \(\bm q=(\bm a-\bm b)/2\)
\item Set \(U=2^{r-1}\) and use the cyclic groups in \eqref{eq:cyclic}
\item Synthesize \(y\) and the \(r\) differential bases \(z_g\)
\item Set \(\beta_g=1\), \(s_0=y+\sum_gz_g\), \(s_1=y-\sum_gz_g\)
\item Set \(u^\star=0\), evaluate \(J^\star=J(\one)\), and store \((s_0,s_1)\)
\item \textbf{for} \(u=1,\ldots,U-1\) in Gray code order
\item \hspace*{0.7em} Identify the changed group sign \(g\)
\item \hspace*{0.7em} Update \((s_0,s_1)\leftarrow(s_0-2\beta_gz_g,\ s_1+2\beta_gz_g)\) and \(\beta_g\leftarrow-\beta_g\)
\item \hspace*{0.7em} Evaluate \(J_u\) from \((s_0,s_1)\) using \eqref{eq:objective}
\item \hspace*{0.7em} \textbf{if} \(J_u<J^\star\), set \(J^\star\leftarrow J_u\), \(u^\star\leftarrow u\), and store \((s_0,s_1)\)
\item \textbf{end for}
\item Return \(u^\star\) and its stored layer waveforms
\end{enumerate}
\end{algorithm}

The cyclic structure also reduces the cost of generating the waveform bases. For equal group sizes, \(M=rG\), the
    spectrum associated with group \(\mathcal I_g\) can be obtained from a \(G\)-point DFT by periodic replication and
    deterministic phase rotations. For general \(M\), the same spectrum follows from an \(M\)-point DFT of the masked
    input \(\bm q\odot\mathbf1_{\mathcal I_g}\). Together with the common waveform, generating all candidates therefore
    requires \(r+1\) \(N\)-point inverse DFTs after subcarrier mapping. Let \(B\) denote this number. Thus \(B=r+1\) and
    \(U=2^{B-2}\), so each additional \(N\)-point inverse DFT doubles the number of available assignments. The fixed
    assignment corresponds to \(B=2\).

After fixing \(\beta_{0,u}=1\), the candidates differ only in the remaining \(r-1\) group signs. Ordering these signs by
    a binary reflected Gray code makes consecutive candidates differ in one group only. If candidates \(u\) and \(v\)
    differ in group \(g\), their layer waveforms are updated as
\[
s_0^{(v)}=s_0^{(u)}-2\beta_{g,u}z_g,
\qquad
s_1^{(v)}=s_1^{(u)}+2\beta_{g,u}z_g.
\]
Hence the waveform bases are generated once, while the candidates are enumerated by simple updates.
    Algorithm~\ref{alg:assignment-selection} summarizes the resulting selection procedure.

The cyclic partitions also support incremental refinement. Consider nested group counts \(r_1<\cdots<r_S\), where each
    finer partition splits every group of the preceding partition. A coarse differential basis is then the sum of its
    finer bases, so moving from \(r_i\) to \(r_{i+1}\) requires only \(r_{i+1}-r_i\) additional \(N\)-point inverse
    DFTs. Let \(J_i\) denote the selected PAPR after stage \(i\). Because the candidate sets are nested, \(J_{i+1}\le J_i\).

With a threshold \(\tau\), refinement stops at the first stage satisfying \(J_i\le\tau\), or at the finest stage
    otherwise. Since stage \(i\) uses \(r_i+1\) \(N\)-point inverse DFTs, the average count is
\[
\bar B=r_1+1+\sum_{i=1}^{S-1}(r_{i+1}-r_i)\Pr(J_i>\tau).
\]
Let \(J_{\rm ref}\) denote the resulting PAPR. For any \(\gamma>\tau\), a block that stops early has \(J_{\rm
    ref}\le\tau<\gamma\), whereas \(J_S>\gamma\) implies \(J_i>\tau\) at every earlier stage. Hence \(\Pr(J_{\rm
    ref}>\gamma)=\Pr(J_S>\gamma)\). Refinement therefore preserves the PAPR complementary cumulative distribution
    function (CCDF) of the finest candidate set above \(\tau\) while reducing basis generation for blocks that stop earlier.

\section{Numerical Results}

\begin{table}[!t]
\caption{Online operations for the \(B=6\) comparison}
\label{tab:budget_complexity}
\centering
\footnotesize
\setlength{\tabcolsep}{3.3pt}
\renewcommand{\arraystretch}{1.08}
\begin{tabular}{@{}lccc@{}}
\toprule
Component & CLASS & oSLM/dSLM & oPTS-w\\
\midrule
Forward DFTs & \(1\!\times\!600+5\!\times\!120\) & \(6\!\times\!600\) & \(2\!\times\!600\)\\
\(N\)-point inverse DFTs & 6 & 6 & 6\\
PAPR evaluations & 32 & 6 & \(2W^2\)\\
\bottomrule
\end{tabular}
\end{table}

\begin{figure}[!t]
\centering
\includegraphics[width=0.80\columnwidth]{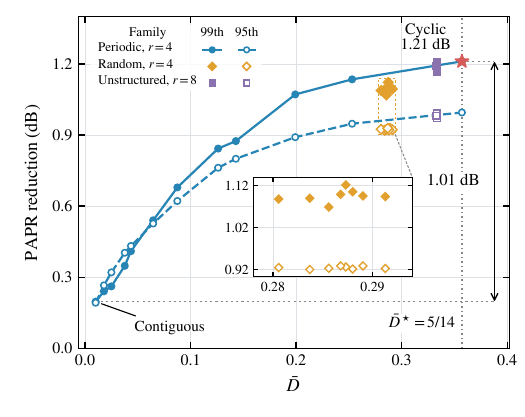}
\caption{PAPR reduction versus candidate diversity at \(U=8\). Cyclic grouping attains the largest \(\bar D\) within the
    family with the minimum basis count \(r=4\) and approaches the reduction of unstructured random assignments with \(r=8\).}
\label{fig:grouping_design}
\end{figure}

We evaluate how the organization of assignment candidates affects PAPR reduction and how CLASS uses the available
    synthesis budget to realize these gains. We consider localized DFT-s-OFDM with \(M=600\), \(N=1024\), and
    independent 16-QAM symbols with unit average energy. Let \(q_{0.01}\) denote the 99th percentile of
    \(10\log_{10}J\). The continuous-time peak in \eqref{eq:objective} is evaluated on 4096 uniformly spaced points over
    one useful symbol, and \(B\) denotes the number of \(N\)-point inverse DFTs used to synthesize a rank-2 block.

The baselines are ordinary SLM (oSLM) and directed SLM (dSLM) \cite{FischerSiegl2009}, and ordinary PTS with phase
    weighting (oPTS-w) \cite{SieglFischer2008}. At budget \(B\), oSLM distributes its trials as evenly as possible
    between the two layers, whereas dSLM directs each additional trial to the layer with the larger current PAPR. For
    oPTS-w, each layer uses \(v\) adjacent subblocks with \(B=2v\) and phase alphabet size \(W\in\{2,4\}\). Unless
    otherwise stated, \(B=6\), giving \(r=5\), \(U=16\), and \(G=120\) for CLASS.

\begin{figure}[!t]
\centering
\includegraphics[width=0.80\columnwidth]{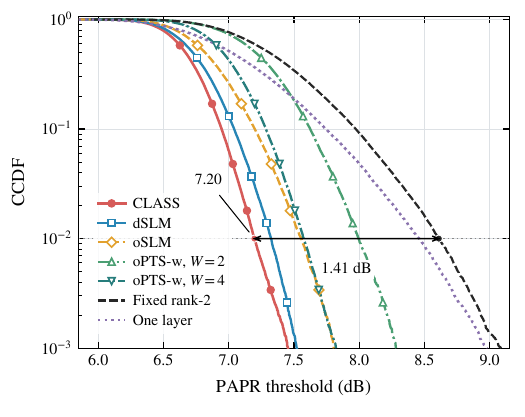}
\caption{PAPR CCDFs at \(B=6\). CLASS reduces the 99th percentile PAPR by 1.41 dB relative to a fixed rank-2 assignment.}
\label{fig:ccdf}
\end{figure}

\subsection{Candidate Design}
At \(U=8\), Fig.~\ref{fig:grouping_design} relates assignment diversity to the resulting PAPR reduction. Within the
    family with the minimum basis count \(r=4\), cyclic grouping attains the theoretical maximum \(\bar D=5/14\) and
    reduces \(q_{0.01}\) by 1.21 dB, whereas contiguous grouping gives only 0.20 dB. The other periodic and balanced
    random partitions follow the same overall trend. Unstructured random assignments with \(r=8\) achieve 1.17--1.21 dB,
    showing that the cyclic construction retains essentially the same reduction using only four differential bases.

\subsection{PAPR Reduction}
We next compare PAPR under the same waveform synthesis budget. At \(B=6\), all methods use six \(N\)-point inverse DFTs
    (Table~\ref{tab:budget_complexity}), while CLASS forms 16 assignments from five differential bases.
    Fig.~\ref{fig:ccdf} shows the resulting CCDFs. At CCDF \(10^{-2}\), CLASS reduces \(q_{0.01}\) from 8.61 dB for a
    fixed rank-2 assignment to 7.20 dB, compared with 7.32 dB for dSLM, 7.56 dB for oSLM, 7.99 dB for oPTS-w with
    \(W=2\), and 7.57 dB with \(W=4\). It therefore gives the lowest 99th percentile PAPR among the compared schemes,
    with gains of 1.41 dB over the fixed rank-2 waveform and 1.26 dB over the 8.46 dB one-layer reference.

\subsection{Synthesis Budget and Selective Refinement}
\begin{figure}[!t]
\centering
\includegraphics[width=0.80\columnwidth]{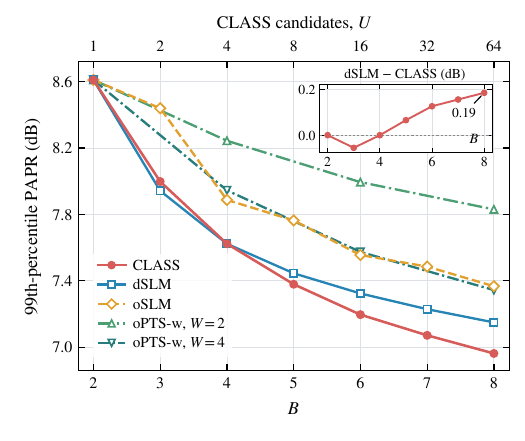}
\caption{99th percentile PAPR versus the number \(B\) of \(N\)-point inverse DFTs per rank-2 block. The top axis gives
    the number of CLASS assignments, and the inset shows dSLM minus CLASS.}
\label{fig:synthesis_budget}
\end{figure}

The benefit of shared waveform bases becomes clearer as the synthesis budget increases. Fig.~\ref{fig:synthesis_budget}
    shows that CLASS is comparable to dSLM for small \(B\) and becomes better from \(B=5\) onward. At \(B=8\), it
    reaches 6.96 dB, compared with 7.15 dB for dSLM, 7.37 dB for oSLM, 7.83 dB for oPTS-w with \(W=2\), and 7.34 dB with
    \(W=4\). The gain over dSLM thus grows to 0.19 dB at the largest tested budget.

Selective refinement can further avoid unnecessary synthesis on blocks whose PAPR is already low. With $\tau_{\rm
    dB}=7.05$ dB, selective refinement with $r=1,3,6$ uses 4.36 $N$-point inverse DFTs on average instead of seven, a
    37.7\% reduction, while retaining $q_{0.01}=7.07$ dB of the complete $r=6$ search.

\section{Conclusion}
This paper developed CLASS, a structured layer-assignment framework for low-PAPR rank-2 DFT-s-OFDM transmission. The
    results reveal that effective PAPR reduction depends not simply on the number of candidate waveforms, but on how
    their diversity is organized relative to the cost of generating them. By exploiting the shared structure among
    candidates, carefully designed assignments can achieve comparable performance to more computationally demanding
    searches with substantially fewer waveform synthesis operations. This suggests a shift in waveform design from
    generating more alternatives to extracting greater benefit from each synthesis operation. Looking ahead, translating
    this approach into practical 6G uplink systems will require accounting for realistic power-amplifier
    characteristics, spatial precoding, and the signaling overhead associated with assignment selection.

\bibliographystyle{IEEEtran}
\bibliography{refs}
\end{document}